\documentclass[12pt]{article}
\usepackage{pic04}
\usepackage{hyperref}
\usepackage{url}
\usepackage{graphicx}

\begin{document}

%
%
\newcommand{\mylist}[2]
  {\begin{list}{#1}{
      \setlength{\topsep}{0 pt}
      \setlength{\parsep}{0.0in}
      \setlength{\itemsep}{#2}
  }}
\newcommand{\newlist}[3]
  {\begin{list}{#1}{
      \setlength{\topsep}{0 pt}
      \setlength{\parsep}{0.0in}
      \setlength{\itemsep}{#2}
      \setlength{\leftmargin}{#3}
  }}
\newcommand{\D}{\displaystyle}
\newcommand{\T}{\textstyle}
\newcommand{\Gam}{\mbox{$\Gamma$}}
\newcommand{\aGam}{\mbox{$\overline{\Gamma}$}}
\newcommand{\CP}{\mbox{\it CP}}
\newcommand{\ra}{\mbox{~$\rightarrow$}~} 
\newcommand{\la}{\mbox{~$\leftarrow$}~} 
\newcommand{\lra}{\mbox{~$\leftrightarrow$}~}
\newcommand{\Llra}{\mbox{~$\Longleftrightarrow$}~}
\newcommand{\ccc}{\mbox{$^\circ$}}
\newcommand{\crc}{\mbox{$^{\circ}$}}
\newcommand{\mss}{\mbox{m/s$^2$}}
\newcommand{\kmc}{\mbox{kg/m$^3$}}
\newcommand{\cd}{\mbox{$\cdot$}}
\newcommand{\inch}{\mbox{$^{\prime\prime}$}}
\newcommand{\ft}{\mbox{$^{\prime}$}}
\newcommand{\GeVc}{\mbox{GeV/$c$}}
\newcommand{\GeVcc}{\mbox{GeV/$c^2$}}
\newcommand{\HyperCP}
           {\mbox{\slshape Hyper}\raisebox{-.2ex}{\slshape C}%
                                 \raisebox{.2ex}{\slshape P}}
%
%
%
\newcommand{\ggl}{\mbox{\it g}}
%
\newcommand{\gq}{\mbox{\it q}}
\newcommand{\uq}{\mbox{\it u}}
\newcommand{\dq}{\mbox{\it d}}
\newcommand{\cq}{\mbox{\it c}}
\newcommand{\sq}{\mbox{\it s}}
\newcommand{\tq}{\mbox{\it t}}
\newcommand{\bq}{\mbox{\it b}}
\newcommand{\agq}{\mbox{$\bar{q}$}}
\newcommand{\auq}{\mbox{$\bar{u}$}}
\newcommand{\adq}{\mbox{$\bar{d}$}}
\newcommand{\acq}{\mbox{$\bar{c}$}}
\newcommand{\asq}{\mbox{$\bar{s}$}}
\newcommand{\atq}{\mbox{$\bar{t}$}}
\newcommand{\abq}{\mbox{$\bar{b}$}}
\newcommand{\X}{\mbox{\it X}}
\newcommand{\Xm}{\mbox{\it X$^-$}}
\newcommand{\Xp}{\mbox{\it X$^+$}}
\newcommand{\Xz}{\mbox{\it X$^{\circ}$}}
%
\newcommand{\gl}{\mbox{\it l}}
\newcommand{\lm}{\mbox{\it l$^-$}}
\newcommand{\lp}{\mbox{\it l$^+$}}
\newcommand{\lpm}{\mbox{\it l$^{\pm}$}}
\newcommand{\lmp}{\mbox{\it l$^{\mp}$}}
\newcommand{\gel}{\mbox{$e$}}
\newcommand{\agel}{\mbox{$\overline{e}$}}
\newcommand{\elp}{\mbox{$e^+$}}
\newcommand{\elm}{\mbox{$e^-$}}
\newcommand{\elpm}{\mbox{$e^{\pm}$}}
\newcommand{\elmp}{\mbox{$e^{\mp}$}}
\newcommand{\gmu}{\mbox{$\mu$}}
\newcommand{\mum}{\mbox{$\mu^-$}}
\newcommand{\mup}{\mbox{$\mu^+$}}
\newcommand{\mupm}{\mbox{$\mu^{\pm}$}}
\newcommand{\mump}{\mbox{$\mu^{\mp}$}}
\newcommand{\taum}{\mbox{$\tau^-$}}
\newcommand{\taup}{\mbox{$\tau^+$}}
\newcommand{\gnu}{\mbox{$\nu$}}
\newcommand{\agnu}{\mbox{$\overline{\nu}$}}
\newcommand{\nul}{\mbox{$\nu_l$}}
\newcommand{\anul}{\mbox{$\overline{\nu}_l$}}
\newcommand{\nue}{\mbox{$\nu_e$}}
\newcommand{\num}{\mbox{$\nu_{\mu}$}}
\newcommand{\nut}{\mbox{$\nu_{\tau}$}}
\newcommand{\anue}{\mbox{$\overline{\nu}_e$}}
\newcommand{\anum}{\mbox{$\overline{\nu}_{\mu}$}}
\newcommand{\anut}{\mbox{$\overline{\nu}_{\tau}$}}
\newcommand{\ggam}{\mbox{$\gamma$}}
\newcommand{\aggam}{\mbox{$\overline{\gamma}$}}
\newcommand{\ab}{\mbox{$b\overline{b}$}}
\newcommand{\sbab}{\mbox{$\sigma_{b\overline{b}}$}}
\newcommand{\sJ}{\mbox{$\sigma_{\psi}$}}
\newcommand{\st}{\mbox{$\sigma_T$}}
\newcommand{\gB}{\mbox{\it B}}
\newcommand{\agB}{\mbox{$\overline{\it B}$}}
\newcommand{\gBz}{\mbox{$B^{\circ}$}}
\newcommand{\agBz}{\mbox{$\overline{B}^{\circ}$}}
\newcommand{\gBp}{\mbox{$B^+$}}
\newcommand{\gBm}{\mbox{$B^-$}}
\newcommand{\Bu}{\mbox{$B_u$}}
\newcommand{\Bupm}{\mbox{$B^{\pm}_u$}}
\newcommand{\Bup}{\mbox{$B^{+}_{u}$}}
\newcommand{\Bum}{\mbox{$B^{-}_{u}$}}
\newcommand{\Bd}{\mbox{$B_d$}}
\newcommand{\aBd}{\mbox{$\overline{B}_d$}}
\newcommand{\Bdz}{\mbox{$B^{\circ}_d$}}
\newcommand{\aBdz}{\mbox{$\overline{B}^{\circ}_d$}}
\newcommand{\Bs}{\mbox{$B_s$}}
\newcommand{\aBs}{\mbox{$\overline{B}_s$}}
\newcommand{\Bsz}{\mbox{$B^{\circ}_s$}}
\newcommand{\aBsz}{\mbox{$\overline{B}^{\circ}_s$}}
\newcommand{\Bc}{\mbox{$B^{\pm}_c$}}
\newcommand{\Bl}{\mbox{$\Lambda_b$}}
\newcommand{\piz}{\mbox{$\pi^{\circ}$}}
\newcommand{\gpi}{\mbox{$\pi$}}
\newcommand{\pip}{\mbox{$\pi^+$}}
\newcommand{\pim}{\mbox{$\pi^-$}}
\newcommand{\pipm}{\mbox{$\pi^{\pm}$}}
\newcommand{\pimp}{\mbox{$\pi^{\mp}$}}
\newcommand{\gK}{\mbox{$K$}}
\newcommand{\gaK}{\mbox{$\overline{K}$}}
\newcommand{\Kp}{\mbox{$K^+$}}
\newcommand{\Km}{\mbox{$K^{-}$}}
\newcommand{\Kpm}{\mbox{$K^{\pm}$}}
\newcommand{\Kep}{\mbox{$K^{\ast{+}}$}}
\newcommand{\Kez}{\mbox{$K^{\ast\circ}$}}
\newcommand{\Kepm}{\mbox{$K^{\ast\pm}$}}
\newcommand{\Kem}{\mbox{$K^{\ast{-}}$}}
\newcommand{\Ke}{\mbox{$K^{\ast\circ}$}}
\newcommand{\Kz}{\mbox{$K^{\circ}$}}
\newcommand{\aKz}{\mbox{$\overline{K}^{\circ}$}}
\newcommand{\aKzr}{\mbox{$\overline{K}^{\circ}_r$}}
\newcommand{\aKzl}{\mbox{$\overline{K}^{\circ}_l$}}
\newcommand{\Ks}{\mbox{$K_{S}$}}
\newcommand{\Ksz}{\mbox{$K_{S}^{\circ}$}}
\newcommand{\Kl}{\mbox{$K_{L}$}}
\newcommand{\Kone}{\mbox{$K_1$}}
\newcommand{\Ktwo}{\mbox{$K_2$}}
\newcommand{\epepp}{\mbox{$\epsilon^{\prime}/\epsilon$}}
\newcommand{\Jpsi}{\mbox{J/$\psi$}}
\newcommand{\psip}{\mbox{$\psi^{\prime}$}}
\newcommand{\Jp}{\mbox{$\psi(2S)$}}
\newcommand{\chione}{\mbox{$\chi_1$}}
\newcommand{\chitwo}{\mbox{$\chi_2$}}
\newcommand{\rhoz}{\mbox{$\rho^{\circ}$}}
\newcommand{\rhop}{\mbox{$\rho^+$}}
\newcommand{\rhom}{\mbox{$\rho^-$}}
\newcommand{\rhopm}{\mbox{$\rho^{\pm}$}}
\newcommand{\etaz}{\mbox{$\eta^{\circ}$}}
\newcommand{\etap}{\mbox{$\eta\prime$}}
\newcommand{\gD}{\mbox{$D$}}
\newcommand{\agD}{\mbox{$\overline{D}$}}
\newcommand{\Dp}{\mbox{$D^+$}}
\newcommand{\Dm}{\mbox{$D^-$}}
\newcommand{\Du}{\mbox{$D_u$}}
\newcommand{\aDu}{\mbox{$\overline{D}_u$}}
\newcommand{\Dstar}{\mbox{$D^{\ast}$}}
\newcommand{\aDstar}{\mbox{$\overline{D}^{\ast}$}}
\newcommand{\Dz}{\mbox{$D^{\circ}$}}
\newcommand{\aDz}{\mbox{$\overline{D}^{\circ}$}}
\newcommand{\Dstarz}{\mbox{$D^{\ast\circ}$}}
\newcommand{\aDstarz}{\mbox{$\overline{D}^{\ast\circ}$}}
\newcommand{\Dstarp}{\mbox{$D^{\ast{+}}$}}
\newcommand{\Dsm}{\mbox{$D^{-}_{s}$}}
\newcommand{\Dsp}{\mbox{$D^{+}_{s}$}}
\newcommand{\Lc}{\mbox{$\Lambda^+_c$}}
\newcommand{\Tu}{\mbox{$T_u$}}
\newcommand{\aTu}{\mbox{$\overline{T}_u$}}
\newcommand{\Tc}{\mbox{$T_c$}}
\newcommand{\aTc}{\mbox{$\overline{T}_c$}}
\newcommand{\rgp}{\mbox{p}}
\newcommand{\ragp}{\mbox{$\overline{\rm p}$}}
\newcommand{\gp}{\mbox{$p$}}
\newcommand{\agp}{\mbox{$\overline{p}$}}
\newcommand{\gppm}{\mbox{$p^{\pm}$}}
\newcommand{\gpmp}{\mbox{$p^{\mp}$}}
\newcommand{\ppm}{\mbox{$p^{\pm}$}}
\newcommand{\pmp}{\mbox{$p^{\mp}$}}
\newcommand{\gn}{\mbox{\it n}}
\newcommand{\agn}{\mbox{$\overline{\it n}$}}
\newcommand{\gL}{\mbox{$\Lambda$}}
\newcommand{\agL}{\mbox{$\overline{\Lambda}$}}
\newcommand{\Lz}{\mbox{$\Lambda^{\circ}$}}
\newcommand{\aLz}{\mbox{$\overline{\Lambda}$$^{\circ}$}}
\newcommand{\gXi}{\mbox{$\Xi$}}
\newcommand{\agXi}{\mbox{$\overline{\Xi}$}}
\newcommand{\Xipm}{\mbox{$\Xi^{\pm}$}}
\newcommand{\Xim}{\mbox{$\Xi^-$}}
\newcommand{\aXim}{\mbox{$\overline{\Xi}$$^+$}}
\newcommand{\Xiz}{\mbox{$\Xi^{\circ}$}}
\newcommand{\aXiz}{\mbox{$\overline{\Xi}$$^{\circ}$}}
\newcommand{\Sigz}{\mbox{$\Sigma^{\circ}$}}
\newcommand{\aSigz}{\mbox{$\overline{\Sigma}$$^{\circ}$}}
\newcommand{\Sigp}{\mbox{$\Sigma^+$}}
\newcommand{\Sigm}{\mbox{$\Sigma^-$}}
\newcommand{\Sigpm}{\mbox{$\Sigma^{\pm}$}}
\newcommand{\gOm}{\mbox{$\Omega$}}
\newcommand{\agOm}{\mbox{$\overline{\Omega}$}}
\newcommand{\Omm}{\mbox{$\Omega^-$}}
\newcommand{\Ompm}{\mbox{$\Omega^{\pm}$}}
\newcommand{\aOmm}{\mbox{$\overline{\Omega}$$^+$}}
%
%
\newcommand{\Xic}{\mbox{${\Xi}^{\circ}_{c}$}}
\newcommand{\aXic}{\mbox{$\overline{\Xi}^{\circ}_{c}$}}
\newcommand{\btoba}{\makebox{$|\langle\aBd|\Bd\rangle|^2$}}
\newcommand{\batob}{\makebox{$|\langle\Bd|\aBd\rangle|^2$}}
\newcommand{\btob}{\makebox{$|\langle\Bd|\Bd\rangle|^2$}}
\newcommand{\batoba}{\makebox{$|\langle\aBd|\aBd\rangle|^2$}}
\newcommand{\Psipm}{\makebox{$\Psi_{\pm}$}}
\newcommand{\Psit}{\makebox{$\Psi(t)$}}
\newcommand{\Psiat}{\makebox{$\bar{\Psi}(t)$}}
\newcommand{\Bzb}{\makebox{$|B^{\circ}\rangle$}}
\newcommand{\Bzab}{\makebox{$|\bar{B}^{\circ}\rangle$}}
%
%
\newcommand{\Pol}{\makebox{$\vec{P}$}}
\newcommand{\PolL}{\makebox{$\vec{P}_{\Lambda}$}}
\newcommand{\PolXi}{\makebox{$\vec{P}_{\Xi}$}}
\newcommand{\Polp}{\makebox{$\vec{P}_p$}}
\newcommand{\Pold}{\makebox{$\vec{P}_d$}}
%
%
\newcommand{\php}{\makebox{$\hat{p}_p$}}
\newcommand{\phd}{\makebox{$\hat{p}_d$}}
\newcommand{\ph}{\makebox{$\hat{p}$}}
\newcommand{\phL}{\makebox{$\hat{p}_{\Lambda}$}}
\newcommand{\phX}{\makebox{$\hat{p}_{\Xi}$}}
%
%
\newcommand{\galpha}{\mbox{$\alpha$}}
\newcommand{\agalpha}{\mbox{$\overline{\alpha}$}}
\newcommand{\gbeta}{\mbox{$\beta$}}
\newcommand{\agbeta}{\mbox{$\overline{\beta}$}}
\newcommand{\ggamma}{\mbox{$\gamma$}}
\newcommand{\aggamma}{\mbox{$\overline{\gamma}$}}
\newcommand{\alp}{\makebox{$\alpha_p$}}
\newcommand{\bp}{\makebox{$\beta_p$}}
\newcommand{\gap}{\makebox{$\gamma_p$}}
\newcommand{\alal}{\makebox{$\alpha\alpha$}}
\newcommand{\alalbar}{\makebox{$\overline{\alpha}\overline{\alpha}$}}
\newcommand{\delalal}{\makebox{$\delta\alpha\alpha$}}
\newcommand{\delalalbar}{\makebox{$\delta\overline{\alpha}\overline{\alpha}$}}
\newcommand{\alXi}{\makebox{$\alpha_{\Xi}$}}
\newcommand{\aalXi}{\makebox{$\alpha_{\overline{\Xi}}$}}
\newcommand{\aalXibig}{\makebox{$\overline{\alpha}_{\Xi}$}}
\newcommand{\bXi}{\makebox{$\beta_{\Xi}$}}
\newcommand{\gaXi}{\makebox{$\gamma_{\Xi}$}}
\newcommand{\gamXi}{\makebox{$\gamma_{\Xi}$}}
\newcommand{\alL}{\makebox{$\alpha_{\Lambda}$}}
\newcommand{\aalL}{\makebox{$\alpha_{\overline{\Lambda}}$}}
\newcommand{\aalLbig}{\makebox{$\overline{\alpha}_{\Lambda}$}}
\newcommand{\alOm}{\makebox{$\alpha_{\Omega}$}}
\newcommand{\aalOm}{\makebox{$\alpha_{\overline{\Omega}}$}}
\newcommand{\aalOmbig}{\makebox{$\overline{\alpha}_{\Omega}$}}
\newcommand{\betaOm}{\makebox{$\beta_{\Omega}$}}
\newcommand{\delaL}{\makebox{$\Delta\alpha_{\Lambda}$}}
\newcommand{\delaXi}{\makebox{$\Delta\alpha_{\Xi}$}}
\newcommand{\AXi}{\makebox{$A_{\Xi}$}}
\newcommand{\aAXi}{\makebox{$A_{\overline{\Xi}}$}}
\newcommand{\Aratio}{\makebox{$\frac{\alpha + \overline{\alpha}}%
                                    {\alpha - \overline{\alpha}}$}}
\newcommand{\Aratiobig}{\makebox{$\D\frac{\alpha + \overline{\alpha}}%
                                         {\alpha - \overline{\alpha}}$}}
\newcommand{\AXiratio}{\makebox{$\frac{\alXi + \aalXi}{\alXi - \aalXi}$}}
\newcommand{\AXiratiobig}{\makebox{$\D\frac{\alXi + \aalXibig}{\alXi - \aalXibig}$}}
\newcommand{\AXiratiotxt}{\makebox{$(\alXi + \aalXi)/(\alXi - \aalXi)$}}
\newcommand{\AXiratiobigtxt}{\makebox{$(\alXi+\aalXibig)/(\alXi-\aalXibig)$}}
\newcommand{\AL}{\makebox{$A_{\Lambda}$}}
\newcommand{\aAL}{\makebox{$A_{\overline{\Lambda}}$}}
\newcommand{\ALratio}{\makebox{$\frac{\alL + \aalL}{\alL - \aalL}$}}
\newcommand{\ALratiobig}{\makebox{$\D\frac{\alL + \aalLbig}{\alL - \aalLbig}$}}
\newcommand{\ALratiotxt}{\makebox{$(\alL + \aalL)/(\alL - \aalL)$}}
\newcommand{\ALratiobigtxt}{\makebox{$(\alL + \aalLbig)/(\alL - \aalLbig)$}}
\newcommand{\AXiL}{\makebox{${A}_{\Xi\Lambda}$}}
\newcommand{\delAXiL}{\makebox{$\Delta{A}_{\Xi\Lambda}$}}
\newcommand{\deldelta}{\makebox{$\Delta\delta$}}
\newcommand{\AXiLratio}{\makebox{$\frac{\alXi\alL - \aalXi\aalL}
                                       {\alXi\alL + \aalXi\aalL}$}}
\newcommand{\AXiLratiobig}{\makebox{$\D\frac{\alXi\alL - \aalXibig\aalLbig}
                                            {\alXi\alL + \aalXibig\aalLbig}$}}
\newcommand{\AXiLratiotxt}{\makebox{$(\alXi\alL-\aalXi\aalL)/
                                     (\alXi\alL+\aalXi\aalL)$}}
\newcommand{\AXiLratiotxtbig}{\makebox{$(\alXi\alL-\aalXibig\aalLbig)/
                                     (\alXi\alL+\aalXibig\aalLbig)$}}
\newcommand{\abar}{\makebox{$\overline{a}$}}
\newcommand{\LeeYang}{\makebox{$\displaystyle
    \Pold = \frac{(\alp + \Polp\cdot\phd)\phd +
                     \bp(\Polp{\times}\phd) +
                     \gap(\phd{\times}(\Polp{\times}\phd))}
                     {(1{+}\alp\Polp\cdot\phd)}$}}
\newcommand{\LeeYangXi}{\makebox{$\displaystyle
    \PolL = \frac{(\alXi + \PolXi\cdot\phL)\phL +
                     \bXi(\PolXi{\times}\phL) +
                     \gaXi(\phL{\times}(\PolXi{\times}\phL))}
                     {(1{+}\alXi\PolXi\cdot\phL)}$}}
\newcommand{\alphaLY}{\makebox{$\displaystyle
    \alpha = \frac{2\mbox{Re}(S^{\ast}P)}{|S|^2 + |P|^2}$}}
\newcommand{\alphaLYt}{\makebox{$\alpha = {2\mbox{Re}(S^{\ast}P)}/(|S|^2 + |P|^2)$}}
\newcommand{\betaLY}{\makebox{$\displaystyle
    \beta  = \frac{2\mbox{Im}(S^{\ast}P)}{|S|^2 + |P|^2}$}}
\newcommand{\gammaLY}{\makebox{$\displaystyle
    \gamma = \frac{|S|^2 - |P|^2}{|S|^2 + |P|^2}$}} 
\newcommand{\hypdk}{\makebox{$\displaystyle
    \frac{dP}{d\Omega} =
    \frac{1}{4\pi}(1 + \alpha\vec{P}_p{\cdot}\hat{p}_d)$}}
\newcommand{\hypdkp}{\makebox{$\displaystyle
    \frac{dP}{d\cos\theta} =
    \frac{1}{2}(1 + \alpha_pP_p\cos\theta$)}}
\newcommand{\hypdkN}{\makebox{$\displaystyle
    \frac{dN}{d\Omega} =
    \frac{N}{4\pi}(1 + \alpha\vec{P}_p{\cdot}\hat{p}_d)$}}
\newcommand{\hypdkXiL}{\makebox{$\displaystyle
    \frac{dP}{d\cos\theta} =
    \frac{1}{2}(1 + \alpha_{\Xi}\alpha_{\Lambda}\cos\theta)$}}
\newcommand{\hypdkXiLN}{\makebox{$\displaystyle
    \frac{dN}{d\cos\theta} =
    \frac{N}{2}(1 + \alpha_{\Xi}\alpha_{\Lambda}\cos\theta)$}}
%
\newcommand{\alphaOm}{\makebox{$\alpha_{\Omega}$}}
\newcommand{\alphaaOm}{\makebox{$\overline{\alpha}$$_{\Omega}$}}
\newcommand{\alphaL}{\makebox{$\alpha_{\Lambda}$}}
\newcommand{\alphaaL}{\makebox{$\overline{\alpha}$$_{\Lambda}$}}
\newcommand{\alphaXi}{\makebox{$\alpha_{\Xi}$}}
\newcommand{\alphaaXi}{\makebox{$\overline{\alpha}$$_{\Xi}$}}
%
%
\def\kpll{K^+ \rightarrow \pi^+ l^+ l^-}
\def\kmll{K^- \rightarrow \pi^- l^+ l^-}
\def\kpmumu{K^+ \rightarrow \pi^+ \mu^+ \mu^-}
\def\kmmumu{K^- \rightarrow \pi^- \mu^+ \mu^-}
\def\kpmmumu{K^{\pm} \rightarrow \pi^{\pm} \mu^{+}\mu^{-}}
\def\kpee{K^+ \rightarrow \pi^+ e^+ e^-}
\def\kp3pi{K^+ \rightarrow \pi^+ \pi^+ \pi^-}
\def\km3pi{K^- \rightarrow \pi^- \pi^- \pi^+}
\def\kpm3pi{K^{\pm} \rightarrow \pi^{\pm} \pi^+ \pi^-}
\def\kpi{K^{\pm}_{\pi 3}}
\def\kmu{K^{\pm}_{\pi \mu \mu}}
\def\rkppmm{\Gamma(K^+ \rightarrow \pi^+\mu^+\mu^-)}
\def\rkmpmm{\Gamma(K^- \rightarrow \pi^-\mu^+\mu^-)}
\def\rkpmpmm{\Gamma(K^{\pm} \rightarrow \pi^{\pm}\mu^+\mu^-)}
\def\rkpall{\Gamma(K^+ \rightarrow all)}
\def\rkmall{\Gamma(K^- \rightarrow all)}
\def\rkpmall{\Gamma(K^{\pm} \rightarrow all)}
\def\delkpm{\Delta(K^{\pm} \rightarrow \pi^{\pm}\mu^+\mu^-)}
\def\Rkppmm{\Gamma(K^+_{\pi\mu\mu})}
\def\Rkmpmm{\Gamma(K^-_{\pi\mu\mu})}
\def\Rkpmpmm{\Gamma(K^{\pm}_{\pi\mu\mu})}
\def\Delkpm{\Delta(K^{\pm}_{\pi\mu\mu})}
\def\kpmu{K^+_{\pi\mu\mu}}
\def\kmmu{K^-_{\pi\mu\mu}}
\def\kpmmu{K^{\pm}_{\pi\mu\mu}}
\def\kppi{K^+_{{\pi}3}}
\def\kmpi{K^-_{{\pi}3}}
\def\kpmpi{K^{\pm}_{{\pi}3}}
\def\nokpmu{N^{obs}_{K^+_{\pi\mu\mu}}}
\def\nokmmu{N^{obs}_{K^-_{\pi\mu\mu}}}
\def\nokpmmu{N^{obs}_{K^{\pm}_{\pi\mu\mu}}}
\def\nokppi{N^{obs}_{K^+_{\pi{3}}}}
\def\nokmpi{N^{obs}_{K^-_{\pi{3}}}}
\def\nokpmpi{N^{obs}_{K^{\pm}_{\pi{3}}}}
\newcommand{\missmassformula}{
      \[
         M^2_{\rm miss} = M^2_K(1 - \frac{p_{\pi}}{p_K}) +
                          m^2_{\pi}(1 - \frac{p_K}{p_{\pi}}) -
                          p_{\pi}p_K\theta^2
      \]
}

\title{\bf THE SEARCH FOR CP VIOLATION IN HYPERON DECAYS}
\author{E. Craig Dukes \\
{\em University of Virginia, Charlottesville, VA 22901, USA}}
\maketitle
%
%
\begin{figure}[h]
\begin{center}
%
%
%
%
\vspace{4.5cm}
\end{center}
\end{figure}

\baselineskip=14.5pt
\begin{abstract}
Searches for experimental manifestations of \CP\ violation
have born much fruit in recent years with the discovery of direct
\CP\ violation and the first evidence of \CP\ violation 
outside of the neutral kaon system.  Nevertheless we still know little
about \CP\ violation:  its origin remains a mystery
and there is little hard evidence that it is the sole province
of the standard model.  Searches for \CP\ violation
in hyperon decays offer promising possibilities as they
are sensitive to certain beyond-the-standard-model sources that are
not probed in other systems.  We report on the status of such searches,
in particular a new result from the Fermilab \HyperCP\ experiment which
has greatly increased the sensitivity over previous measurements
and is confronting some beyond-the-standard-model theory predictions.
\end{abstract}
\newpage

\baselineskip=17pt

\section{Introduction}

After years of little progress in searching for new manifestations
of \CP\ violation, recently great advances have been made,
with the unambiguous observation of direct-\CP\ violation in kaon decays
\cite{direct_k} and the first observation of \CP\ violation outside
of the neutral kaon system: in decays of the \Bd\ meson
\cite{b_cp}.  Nevertheless, our fundamental understanding of \CP\
violation has improved little in the forty years since its
discovery.  Although \CP\ violation is accommodated quite 
nicely in the standard model --- in the complex phase of the CKM matrix ---
its origin remains a mystery.  And although \CP\ violation is
expected to be ubiquitous in weak interactions, albeit often
vanishingly small, the experimental evidence is still meager.
In addition, many beyond-the-standard-model theories
can produce relatively large \CP-violating effects, none of which
have yet been seen.
It behooves us then to search for other manifestations of this
phenomenon.
Hyperon decays offer a promising venue for such searches
as hyperons can be copiously produced, are easily detected in
experiments of modest cost, and are particularly
sensitive to certain exotic sources of \CP\ violation.  
We review here the status of such searches.

\section{Signatures for \protect\textit{\protect\textbf{CP}} Violation in Hyperon Decays}

The most accessible signature for \CP\ violation in spin-1/2 hyperons
is the comparison of the angular decay distribution of the daughter
baryon with that of the conjugate antibaryon in their two-body nonleptonic
weak decays.  These distributions are not isotropic
because of parity violation, but are given by:
\begin{equation}
   \frac{dN}{d\cos\theta} = \frac{N_0}{2}(1 + \alpha{P_p}\cos\theta),
\label{eq:1}
\end{equation}
where $P_p$ is the parent hyperon polarization, $\cos\theta$ is the daughter
baryon direction in the rest frame of the parent,
and $\alpha = 2\mbox{Re}(S^{\ast}P)/(|S|^2 + |P|^2)$
where $S$ and $P$ are the usual angular momentum amplitudes.

The behavior of the $\alpha$ parameter under \CP\ is 
illustrated for the \gL\ra\gp\pim\ decay in Fig.~\ref{fig:cp_lambda_dk}.  
The daughter proton is emitted preferentially in the
direction of the \gL\ polarization.  Under \CP\ the opposite is
true:  the daughter antiproton is preferentially emitted in the
opposite direction of the \agL\ polarization.  If \CP\ is
good $\agalpha = -\galpha$; hence a difference in the magnitudes
of the hyperon and antihyperon alpha parameters, or equivalently,
their angular decay distributions, is evidence of \CP\ violation.
\begin{figure}[htb]
\centerline{\includegraphics[height=1.5in]{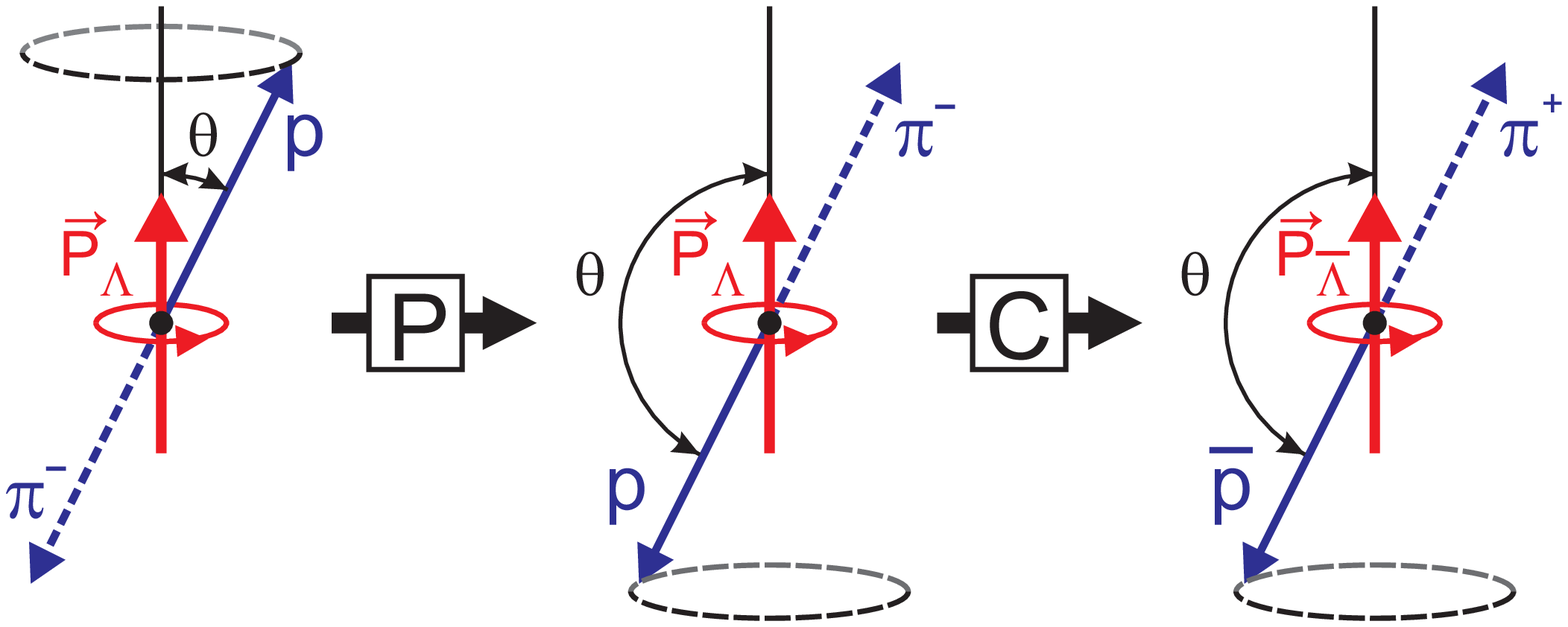}}
\caption{The \gL\ra\gp\pim\ decay under $P$ and $C$ transformations.
  \label{fig:cp_lambda_dk}}
\end{figure}
To extract \galpha, hyperons whose polarizations
are {\em exactly known} are needed, as measuring the daughter
baryon $\cos\theta$ distribution via Eq.~(\ref{eq:1}) gives $\alpha{P}_p$.
Several different methods of producing known polarizations are discussed below.
Hyperons with large $\alpha$ parameters, large production cross
sections, and all charged decay products are favored for
obvious experimental reasons.   Therefore the focus has been on 
\gL\ra\gp\pim\ and \Xim\ra\gL\pim\ra\gp\pim\pim\ (and their conjugate) decays.

\section{Theoretical Expectations}

For \CP\ violation to manifest itself three conditions need be met:
there must be at least two different final states, these states
must have nonzero strong final-state phase-shift differences, and different
\CP-violating weak phase shifts.
Model independent expressions for the difference in alpha parameters
have been explicitly calculated for various hyperon decays \cite{JDonoghue3}.
To leading order they are for \gL\ra\gp\pim\ and \Xim\ra\gL\pim\ decays:
\begin{equation}
   A \equiv \Aratiobig  \cong  
                -\tan(\delta_P-\delta_S)\sin(\phi_P-\phi_S),
\end{equation}
where the $\delta$ are the strong phase shifts
and the $\phi$ are the weak phases.
The \CP\ asymmetry results from the interference of {\em S}- and
{\em P}-wave amplitudes, not isospin amplitudes as is the case
in neutral kaon decays.
The strong final-state phase-shift differences
are small --- $7^{\circ}{\pm}1^{\circ}$ for \gp\gpi\ \cite{RoperL65}
and $4.6^{\circ}{\pm}1.8^{\circ}$ for \gL\gpi\ \cite{HuangM04} ---
greatly reducing the size of any potential asymmetry.

A recent standard model calculation of the \CP\ asymmetries 
has values that range from 
$-0.3{\times}10^{-4}{\leq}\AL{\leq}0.4{\times}10^{-4}$
and $-0.2{\times}10^{-4}{\leq}\AXi{\leq}0.1{\times}10^{-4}$ \cite{TandeanJ03}.
These magnitudes are too small to be experimentally observable at
the present time.
However, beyond-the-standard-model theories can produce larger asymmetries
that {\em are not well constrained by kaon \CP\ measurements}
because hyperon \CP\ violation probes both parity-conserving
and parity-violating amplitudes whereas $\epsilon$ and $\epsilon^{\prime}$
probe only parity-violating amplitudes.
For example, a recent paper shows that the
upper bound on the combined asymmetry $\AXiL\equiv\AXi + \AL$ 
from $\epsilon$ and $\epsilon^{\prime}$ measurements is 
${\sim}100{\times}10^{-4}$ \cite{TandeanJ04}.
The supersymmetric calculation of Ref.~\cite{HeX00}, which does not
contribute to $\epsilon^{\prime}$, can produce a value of
\AL\ of $\mathcal{O}(10^{-3})$.  Other beyond-the-standard-model
theories, such as Left-Right mixing models \cite{ChangD95}
also have enhanced \CP\ asymmetries.  Therefore, any observed effect will
almost certainly be due to new physics.

\section{Early Experimental Results}

Three ingredients are needed to mount an experiment to search
for \CP\ violation in hyperon decays.  First, one needs to
produce hyperons and antihyperons whose polarizations are known 
to the level of the desired sensitivity in $A$, 
or at least are known by some symmetry to be equal.  
Second, one needs a large number of events and large hyperon polarizations,
as the error in $A$ is $\sigma = \sqrt{3/(2N\alpha^2P^2)}$,
where $N$ is the number of hyperons and the number of antihyperons.  
Hence, for modest-sized alpha parameters and polarizations, on the order
of 1 billion events are needed for a $10^{-4}$ measurement.
Finally, the systematics must be controlled to the level of
the measurement.

The results of the four experiments that have published searches 
for \CP\ violation in hyperon decays are listed in 
Table~\ref{tab:limits} below.  
Each of the four experiments used a different technique,
all were limited by statistical, not systematic errors,
and none of the experiments was a dedicated hyperon
\CP\ violation experiment.
Three of the four experiments quote results on $A_{\Lambda}$,
the fourth, E756, measured the combined asymmetry \AXiL.
All reported null results: not surprising as none of the experiments has 
penetrated beyond a $10^{-2}$ sensitivity.

\begin{table}[ht]
\caption{Hyperon \CP\ violation searches.\label{tab:limits}}
\begin{center}
\begin{tabular}{cccr@{.}l@{$\pm$}r@{.}lrc}
\hline
\hline
        \multicolumn{1}{c}{Experiment}
                  & \multicolumn{1}{c}{Mode}
                  & \multicolumn{1}{c}{Technique}
                  & \multicolumn{4}{c}{Result}
                  & \multicolumn{1}{c}{Events}
                  & \multicolumn{1}{c}{Date} \\
\hline
R608               & $A_{\Lambda}$      
                   &  \gp\gp\ra\gL{X}, \agp\gp\ra\agL{X} \rule{0.0in}{0.15in}
                   & $-0$ & 02 & 0 & 14
                   & 26\,581
                   & 1985 \\ 
DM2                & $A_{\Lambda}$      
                   & \elp\elm\ra\Jpsi\ra\gL\agL
                   & $+0$ & 01 & 0 & 10 
                   & 770
                   & 1988  \\
PS185              & $A_{\Lambda}$
                   & \gp\agp\ra\gL\agL
                   & $+0$ & 013 & 0 & 022 
                   & 95\,832
                   & 1996  \\
E756               & $A_{\Xi\Lambda}$
                   & \gp{N} \ra \gXi{X}
                   & $+0$ & 012 & 0 & 014
                   & 280\,000
                   & 2000  \\
\hline
\hline
\end{tabular}
\end{center}
\end{table}

The first result in Table~\ref{tab:limits} is from an ISR
experiment (R608) which produced \gL\ and \agL's in
\gp\gp\ra\gL{X} and \agp\gp\ra\agL{X} reactions \cite{R608}.
They find $\alpha{P}(\agL)/\alpha{P}(\gL) = -1.04\pm0.29$ from
17\,028 \gL\ and 9\,553 \agL\ decays.  Assuming
that the \gL\ and \agL\ polarizations are identical --- which is
rigorously true only if the \gL\ and \agL\ polarizations are
independent of the target identity --- their result can be
converted to a measurement of $A_{\Lambda} = 0.02\pm0.14$.

The next two results of Table~\ref{tab:limits} come from
experiments that produced \gL\agL\ pairs exclusively and
hence reduced potential systematic errors due to any temporal
changes in their detectors.
The second result is from
\Jpsi \ra \gL\agL\ decays measured by the
DM2 detector at the Orsay DCI \elp\elm\ colliding ring \cite{DM2}.
Measuring the correlation between the proton and antiproton momenta
allows the product \alL\aalLbig\ to be extracted.
Fixing \alL\ to its known value ($0.642{\pm}0.013$ \cite{PDG}) 
they obtained from 770 events
\aalLbig = $-0.63\pm0.13$ corresponding to $A_{\Lambda} = 0.01\pm0.10$.
The problem with pushing this method further is obvious:  
since only the product \alL\aalLbig\ is measured,
an independent measurement is needed to decouple the two 
to extract the \CP\ asymmetry.

The third result is from a fixed-target experiment at the CERN Low-Energy
Antiproton Ring (LEAR), PS185 \cite{PS185}.
The experiment was designed to have high-acceptance for \gL\agL\ pairs
produced near threshold.  Antiprotons from the LEAR ring impacted
a CH$_{\rm 2}$ target with the decay products from the
exclusive reaction \agp\gp\ra\gL\agL\ momentum
analyzed in a magnetic spectrometer.
The \gL\ and \agL\ were found to have large predominately negative
polarizations which varied from approximately $+0.2$ to $-0.6$.
The two polarizations are rigorously equal
by $C$-parity conservation in strong interactions:  hence knowledge of 
the magnitudes of the polarizations were not needed for the
determination of $A_{\Lambda}$.
The collaboration took data at several different beam momenta,
below and above the \agp\gp\ra\Sigz\gL\ threshold of 1.653~GeV/$c$.
Their last result, from an analysis of part of their entire
dataset, is $A_{\Lambda} = -0.013\pm0.022$.

These early results showed much promise and led to proposals in 
the early 1990s for much higher-statistics experiments, both at 
\elp\elm\ and \gp\agp\ colliders.  The limitations of the DM2 
technique in measuring \AL\ were to be overcome by analyzing 
\Jpsi\ra\gL\agL\ decays produced from {\em polarized} \elp\elm\ collisions
at a tau-charm factory \cite{taucharm}.
There was also much interest in a gas-jet target experiment at 
an upgraded SuperLEAR at CERN \cite{SuperLEAR}.
Unfortunately funding was not forthcoming for either of these projects.
At the same time it became apparent that a fresh approach taking
advantage of the much larger fluxes of hyperons available in a fixed-target
experiment was feasible.
The conventional wisdom had been that any fixed-target \CP\
violation experiment would be impossible due to the problem
of producing hyperons and antihyperons of precisely known polarizations.
However, a group from Berkeley and the University of Virginia
realized a simple solution to the problem (described below) 
which would allow \AXiL\ to be measured with great precision.
This work led to the \HyperCP\ experiment at Fermilab \cite{hypercp_loi}.
The feasibility of this new experimental technique
was successfully tested on data from Fermilab experiment E756
(designed to measure the \Omm\ magnetic moment), which
had run in the late 1980s, resulting in the most sensitive
limit on \CP\ violation in hyperon decays \cite{E756}.

\section{The First Dedicated Hyperon \textit{\textbf{CP}} Violation Experiment: \protect\HyperCP}

The \HyperCP\ experiment produced \gL's and \agL's with {\em almost}
precisely known polarizations by requiring that they come from
\Xim\ra\gL\pim\ and \aXim\ra\agL\pip\ decays.
The \Xim\ and \aXim\ hyperons were required by parity conservation
in the strong interaction to have {\em zero}
polarization by producing them with an average angle of $0^{\circ}$
with respect to the incident proton beam.
A \gL\ from the weak decay of an unpolarized \gXi\ is found
in a pure helicity state with a polarization magnitude given by the 
parent $\Xi$ alpha parameter: $P_{\Lambda} = \alpha_{\Xi} = -0.458$ \cite{PDG}.  
Hence, if \CP\ is a good symmetry in \gXi\ decays, then the $\Lambda$ and 
$\overline{\Lambda}$ have equal and opposite polarizations.
The decay distributions of the proton and antiproton in the frame
in which the $\Lambda$ direction in the \gXi\ rest frame defines 
the polar axis --- the Lambda Helicity Frame in Fig.~\ref{fig:helicity_frames} 
--- are given by:
\begin{equation}
    \frac{dN}{d\cos\theta}
    = \frac{N_0}{2}(1 + \alpha_{\Lambda}{P_{\Lambda}}\cos\theta)
    = \frac{N_0}{2}(1 + \alpha_{\Lambda}\alpha_{\Xi}\cos\theta).
\label{eq:dist}
\end{equation}
If \CP\ symmetry is good in {\em both} \gXi\ and \gL\ decays
then $\overline{\alpha}_{\Xi} = -\alpha_{\Xi}$
and $\overline{\alpha}_{\Lambda} = -\alpha_{\Lambda}$ and the decay
distributions of the proton and antiproton are identical.
Any difference is evidence of \CP\ violation.
It is evident from Eq.~(\ref{eq:dist}) that differences between the slopes
of the two $\cos\theta$ distributions can be due to \CP\
violation in either the $\Xi$ or $\Lambda$ decay; the experiment is 
sensitive to \CP\ violation in both:
\begin{equation}
   A_{\Xi\Lambda}   \equiv A_{\Lambda} + A_{\Xi}
       \cong \frac{\alpha_{\Lambda}\alpha_{\Xi}
                    - \overline{\alpha}_{\Lambda}\overline{\alpha}_{\Xi}}
                   {\alpha_{\Lambda}\alpha_{\Xi}
                    + \overline{\alpha}_{\Lambda}\overline{\alpha}_{\Xi}}.
\end{equation}
\begin{figure}[htb]
\begin{minipage}[t]{2.0in}
\centerline{\includegraphics[width=1.75in]{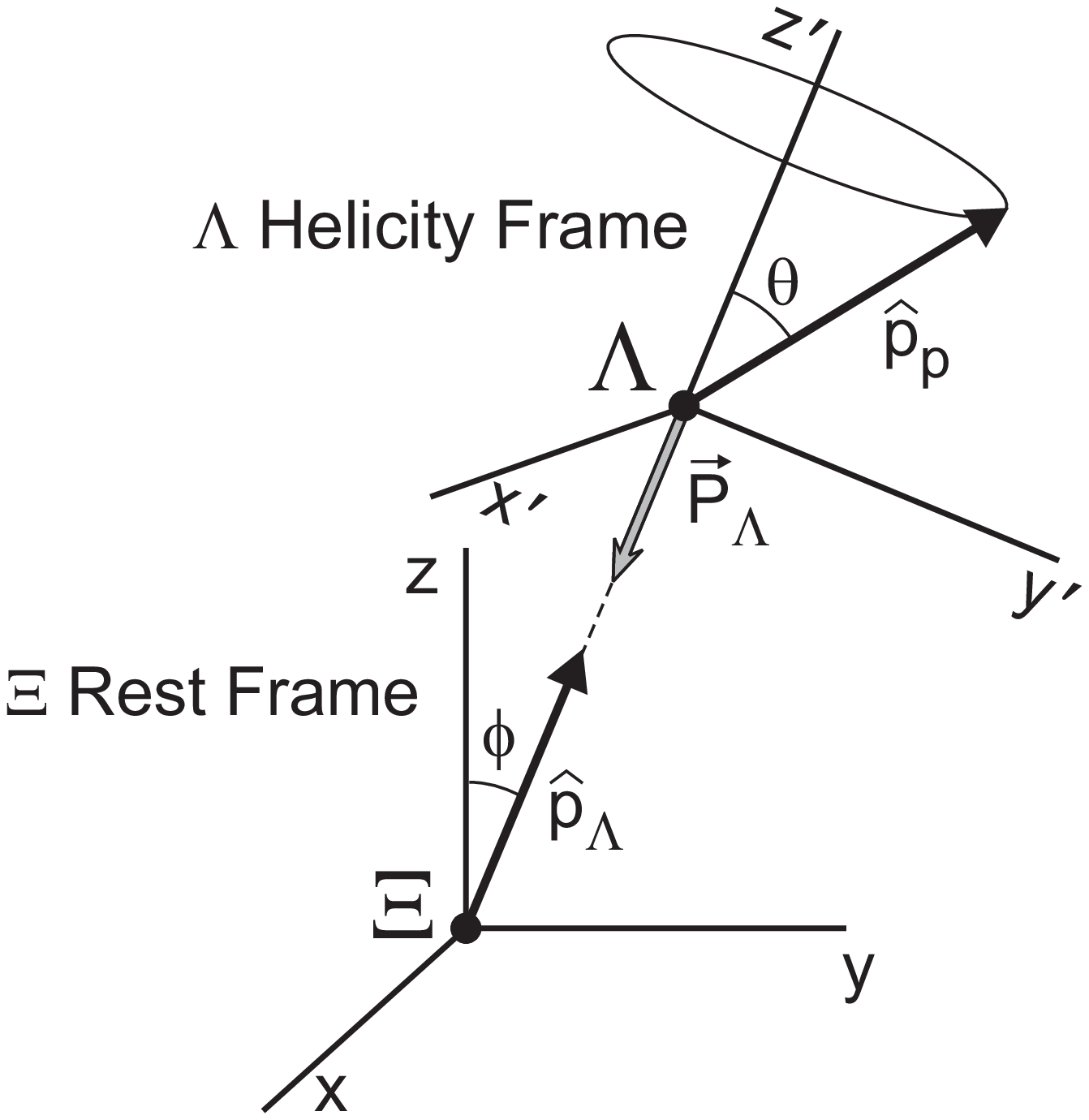}}
\caption{The Lambda Helicity Frame.}
\label{fig:helicity_frames}
\end{minipage}
\hfill
\begin{minipage}[t]{4.0in}
\centerline{\includegraphics[width=3.75in]{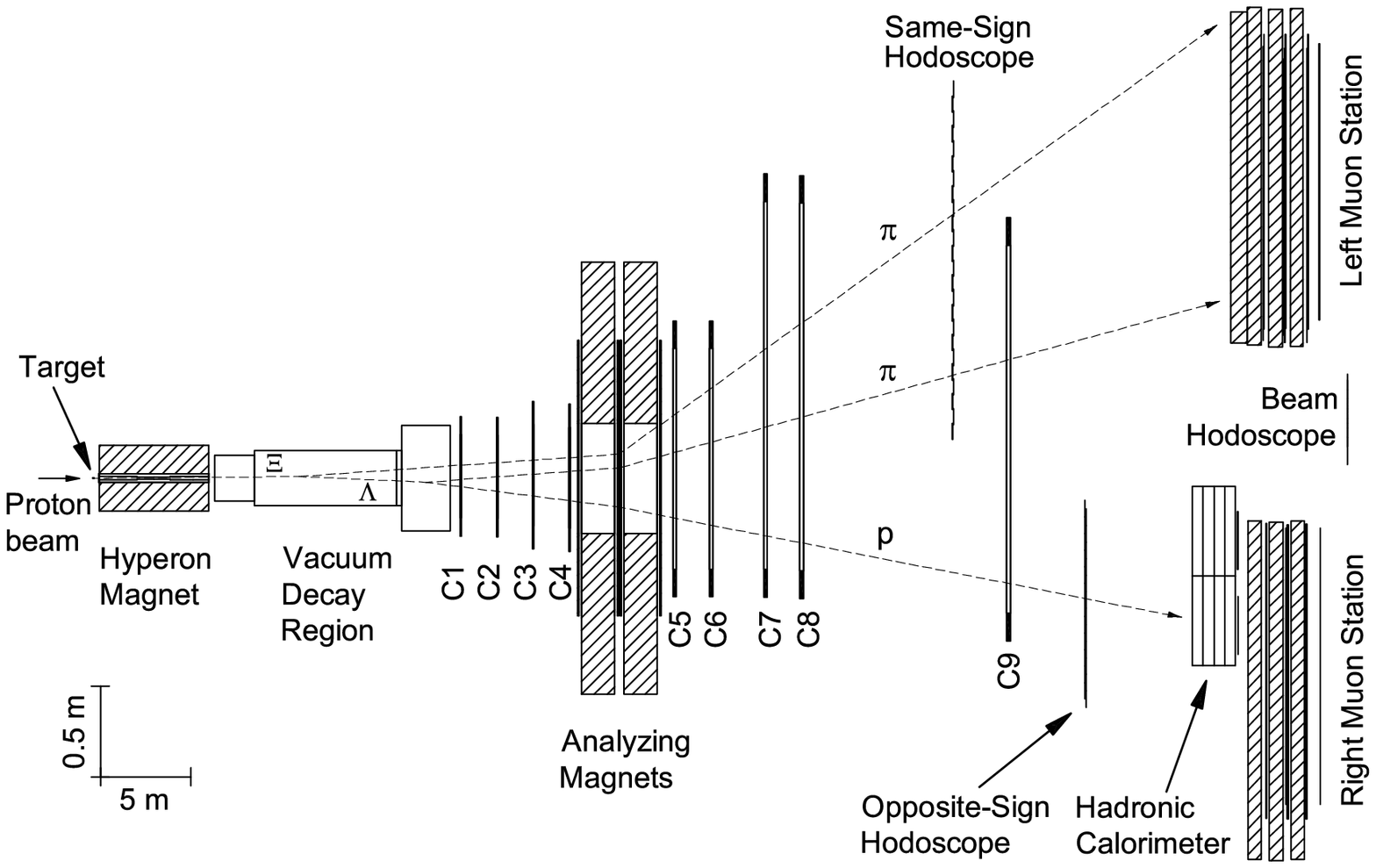}}
\caption{Plan view of the \protect\HyperCP\ apparatus.}
\label{fig:hypercp_plan}
\end{minipage}
\end{figure}

The \HyperCP\ spectrometer (Fig.~\ref{fig:hypercp_plan}) was designed to be
simple, fast, and to have considerable redundancy \cite{spect}.
A charged secondary beam with a mean momentum of about 160\,GeV/$c$ was
produced by steering the Tevatron 800\,GeV/$c$ primary proton beam onto a
$2{\times}2$\,mm$^2$ Cu target which was immediately followed by a 
collimator --- with a 5.38\,$\mu$sr solid angle acceptance --- embedded 
in a 6.1\,m long dipole magnet (Hyperon Magnet).  The central orbit of 
the secondary beam exited the collimator upward at 19.51\,mrad.
Following an evacuated decay region was a magnetic spectrometer 
employing nine high-rate, narrow-pitch wire chambers.  
The spectrometer magnets (Analyzing Magnets) had sufficient field 
integrals to insure that the protons from \gXi\ra\gL\gpi\ra\gp\gpi\gpi\
decays were always deflected to one side of the spectrometer, 
with the two pions deflected to the opposite side, 
and that both were well separated from the intense 
(${\sim}13{\times}10^6$\,s$^{-1}$) secondary beam exiting the collimator.  
A simple trigger was formed by requiring the 
coincidence at the rear of the spectrometer of charged particles  
in two hodoscopes (Same Sign and Opposite Sign Hodoscopes)
situated on either side of the spectrometer,
as well as a minimum amount of energy in a
hadronic calorimeter on the proton side of the spectrometer.
The calorimeter made the trigger ``blind'' to muons and reduced
the trigger rate due to interactions of the secondary beam with
material in the spectrometer. 
The \Xim\ and \aXim\ hyperons were not produced
simultaneously, as is the case with the DM2 and PS185 experiments.
Rather the experiment periodically switched from one running mode
to the other by reversing the polarities of the Hyperon and Analyzing Magnets.

A high-rate data acquisition system enabled up to 100\,000 events
per spill second to be recorded onto magnetic tape.
In two running periods (1997 and 1999) of about 12 months duration
one of the largest data samples ever was recorded, at 231 billion events, 
and by far the largest number of hyperons.
The final dataset was approximately 2.5 billion \Xim\ra\gL\pim\ra\gp\pim\pim\
and \aXim\ra\agL\pip\ra\agp\pip\pip\ decays,
{\em four orders of magnitude more than that of all other hyperon
\CP\ violation searches combined}.  

Since the experiment was probing sensitivities far beyond any 
previously attained, two \CP\ analyses were attempted in parallel.
One analysis separately extracted \alXi\alL\ and \aalXibig\aalLbig\ by 
correcting for the 
apparatus acceptance using a hybrid Monte Carlo (HMC) technique.
The HMC took real \gXi\ra\gL\gpi\ra\gp\gpi\gpi\ events, discarded
the proton and pion, substituting MC-generated proton and pions 
in order to measure the acceptance.  Ten accepted HMC events were 
generated for every real event.  Although this method had the advantage
of providing an absolute measurement of the product of the alpha
parameters, it required an extremely 
faithful MC simulation of the apparatus.  The MC also had to be
extremely fast as on the order of tens of billions of events needed 
to be generated for the analysis of the full dataset.
Although considerable progress was made in refining this
method, it was ultimately abandoned because of systematic difficulties.

The second analysis method was simple:  compare the
proton and antiproton $\cos\theta$ distributions directly, 
without acceptance corrections.
Before this could be done the momentum and spatial distributions of
the \Xim\ and \aXim\ events at the collimator exit (their
effective production point) had to be made identical,
since different production dynamics give different momentum
spectra for the two.  This was done by weighting the
\Xim\ and \aXim\ events in each of the three momentum-dependent parameters
at the collimator exit:  
the magnitude of the \gXi\ momentum, the $y$ slope of the \gXi, and 
the $y$ position of the \gXi.  Each parameter was binned in 
100 bins for a total of one million weights.  
The ratio of the weighted proton and antiproton $\cos\theta$ distributions
was then made.  Any nonzero slope in that ratio is evidence of \CP\ violation.
The ratio was fit to the following form,
\begin{equation}
  R = C\frac{1 + \alXi\alL\cos\theta}
             {1 + (\alXi\alL - \delta)\cos\theta},
\label{eq:fit}
\end{equation}
to extract the asymmetry
$\delta \equiv \alXi\alL - \aalXibig\aalLbig \cong 2\alXi\alL{\cdot}\AXiL$,
where the known value of $\alXi\alL = 0.294$ \cite{PDG} was used.

About 117 (41) million \Xim\ (\aXim) decays selected from the end of
the 1999 run were used --- roughly 10\% of the total dataset. 
Figure~\ref{fig:masses} shows the \Xim\ and \aXim\ masses after all
cuts.  The background under the peak is 0.42\% for both.
The data were divided into 18 parts (Analysis Sets) each of roughly equal size.
Each set was analyzed separately.
Figure~\ref{fig:as1fit} shows the $\cos\theta$ ratios for one of the
Analysis Sets, before and after weighting.
Fits to Eq.~(\ref{eq:fit}) were good: the average chi-squared per 
degree of freedom for all 18 Analysis Sets was 0.97.
\begin{figure}[htb]
\begin{minipage}[t]{2.8in}
\centerline{\includegraphics[width=2.6in]{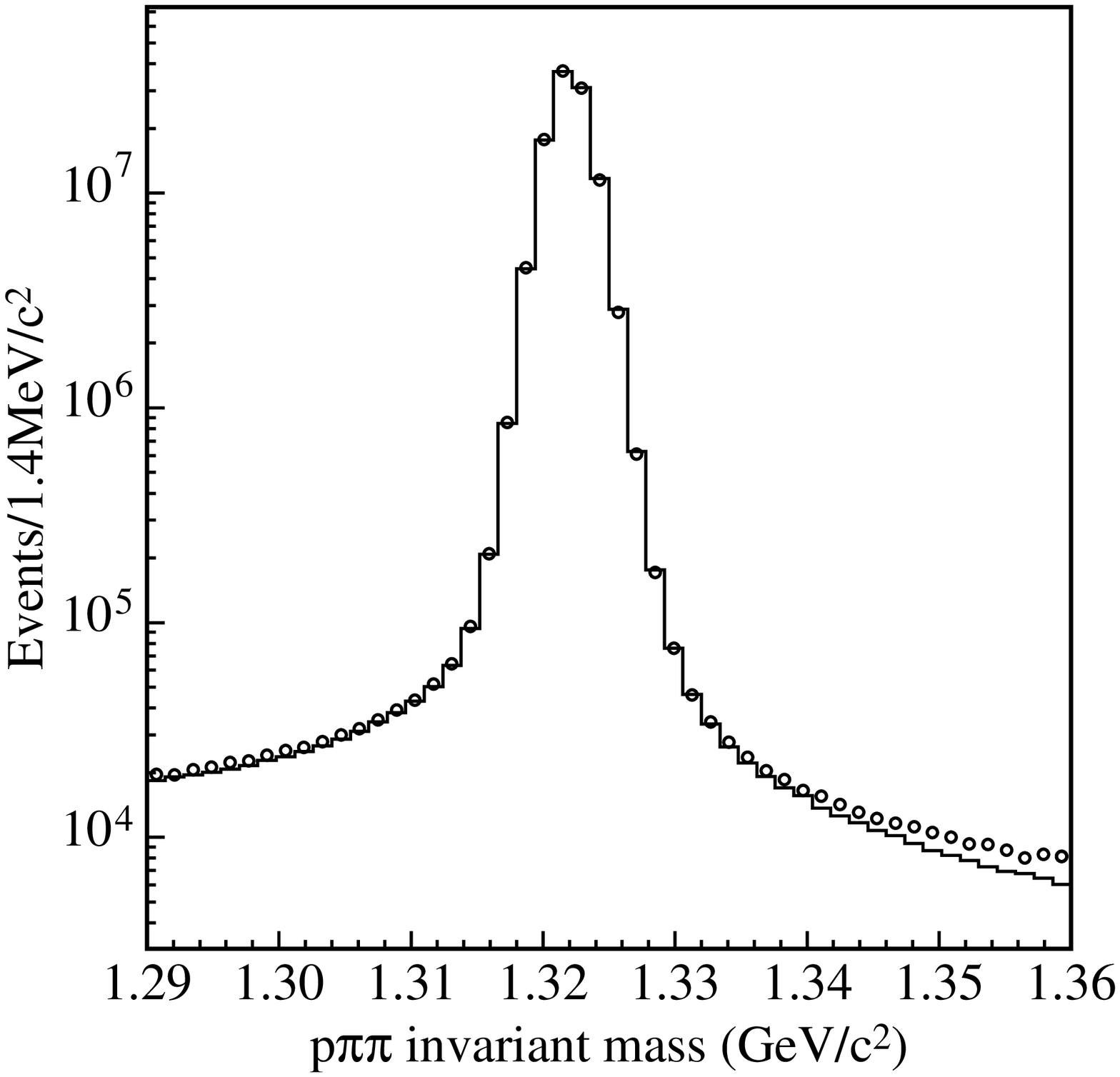}}
\caption{The unweighted \gp\pim\pim\ (histogram) and \agp\pip\pip\
         (circles) invariant masses.
           \label{fig:masses}}
\end{minipage}
\hfill
\begin{minipage}[t]{2.8in}
\centerline{\includegraphics[width=2.7in]{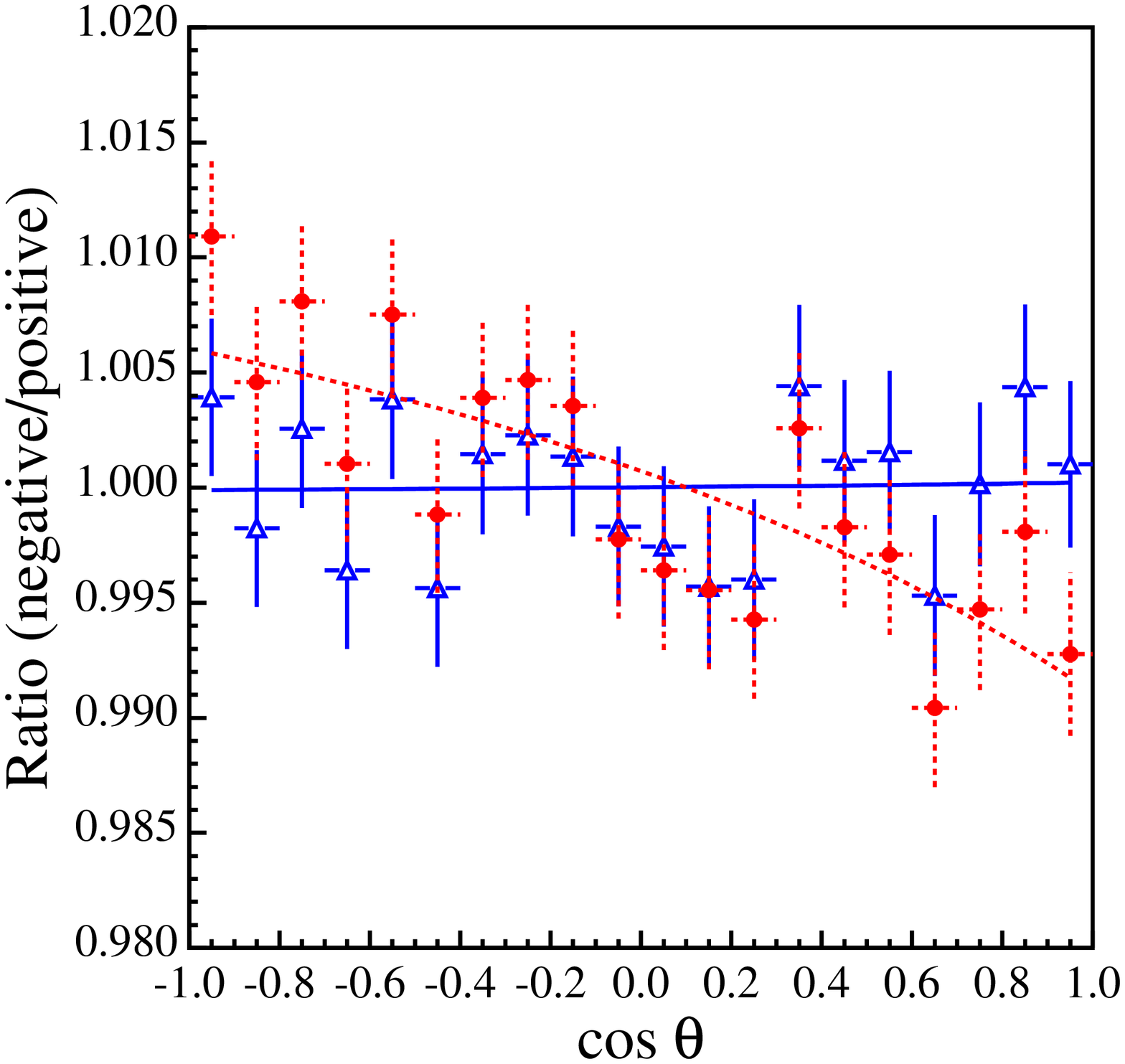}}
\caption{Fits to the weighted (triangles) and unweighted (circles)
         \gp\ to \agp\ $\cos\theta$ ratios from Analysis Set 1.
           \label{fig:as1fit}}
\end{minipage}
\end{figure}

The average asymmetry from all 18 Analysis Sets, after background subtraction 
and with no acceptance or efficiency corrections, was found to be zero:
$\AXiL = [0.0{\pm}5.1({\rm stat}){\pm}4.4({\rm syst})]{\times}10^{-4}$,
with $\chi^2 = 24$.
This is a factor of twenty improvement in sensitivity over the best previous 
result \cite{E756}.
A HMC analysis of about 5\% of the total data sample,
from all of the good 1997 and 1999 runs, also found no evidence of 
a \CP\ asymmetry:
$\AXiL = [-7{\pm}12({\rm stat}){\pm}6({\rm syst})]{\times}10^{-4}$.

Systematic errors were small for several reasons.  First, taking
the ratio of $\cos\theta$ distributions reduced those common to 
the proton and antiproton.  Second, the analysis locked in to the
signal, in a manner analogous to a lock-in amplifier, by
measuring the proton $\cos\theta$ distribution in the Lambda Helicity Frame,
the polar axis of which changed from event to event.
The largest systematic error ($2.4{\times}10^{-4}$) is due to the 
uncertainty in the calibration of the Hall probes situated in the 
Analyzing Magnets.
The next largest ($2.1{\times}10^{-4}$) is the statistics-limited 
uncertainty due to differences in the calorimeter 
efficiencies between positive- and negative-polarity running.
The only other significant systematic error is the uncertainty in
the validation of the analysis code ($1.9{\times}10^{-4}$), again
a statistics-limited result.
Wire chamber and hodoscope efficiency differences were so small that
they were not corrected for, but rather added in as 
negligibly small systematic errors.
No dependence of the asymmetry on \gXi\ momentum, secondary-beam intensity,
or time was found.

The analysis of the entire 1999 \HyperCP\ dataset is well underway and it
is hoped that within a year a result with an improvement in precision
of at least two will be obtained, both in statistical and systematic errors.

\section{Acknowledgments}
I wish to thank the organizers of the XXIV \textsc{Physics in Collision}
for a most enlightening and entertaining conference, and, as always, 
my \HyperCP\ colleagues.  This work was supported
in part by the U.S.\ Department of Energy and the Institute for Nuclear
and Particle Physics at the University of Virginia.

\end{document}